\documentclass[12pt,preprint]{emulateapj}

\usepackage{natbib}
\usepackage{graphicx}
\usepackage{multirow}
\usepackage{txfonts}

\shorttitle{Accretion Dynamics of MAXI~J1659-152 with TCAF}
\shortauthors{D. Debnath, A. A. Molla, S. K. Chakrabarti, $\&$ S. Mondal}

\begin{document}

\title{Accretion Flow Dynamics of MAXI J1659-152 from the Spectral Evolution Study of its 2010 Outburst using the TCAF Solution}

\author{Dipak Debnath\altaffilmark{1}, Aslam Ali Molla\altaffilmark{1}, Sandip K. Chakrabarti\altaffilmark{2,1}, Santanu Mondal\altaffilmark{1}}
\altaffiltext{1}{Indian Center for Space Physics, 43 Chalantika, Garia St. Rd., Kolkata, 700084, India.}
\altaffiltext{2}{S. N. Bose National Centre for Basic Sciences, Salt Lake, Kolkata, 700098, India.}

\email{dipak@csp.res.in; aslam@csp.res.in; chakraba@bose.res.in; santanu@csp.res.in}

\date{Accepted: 10th Feb. 2015; Received: 8th Dec. 2014}

\begin{abstract}

Transient black hole candidates are interesting objects to study in X-rays as these sources 
show rapid evolutions in their spectral and temporal properties. In this paper, we study the spectral properties 
of the Galactic transient X-ray binary MAXI~J1659-152 during its very first outburst after discovery 
with the archival data of RXTE Proportional Counter Array instruments. We make a detailed study of the evolution of accretion 
flow dynamics during its 2010 outburst through spectral analysis using the Chakrabarti-Titarchuk two-component 
advective flow (TCAF) model as an additive table model in XSPEC. 
Accretion flow parameters (Keplerian disk and sub-Keplerian halo rates, 
shock location and shock strength) are extracted from our spectral fits with TCAF.  
We studied variations of these fit parameters during the entire outburst as it passed through 
three spectral classes: {\it hard, hard-intermediate,} and {\it soft-intermediate}. 
We compared our TCAF fitted results with standard combined disk black body (DBB) and power-law (PL) model 
fitted results and found that variations of disk rate 
with DBB flux and halo rate with PL flux are generally similar in nature.
There appears to be an absence of the soft state unlike what is seen in other similar sources. 

\end{abstract}


\section{Introduction}

Compact objects, such as black holes (BHs) and neutron stars are identified by electromagnetic radiations 
emitted from the accreting disk formed by matter from their companion. This matter is accreted either
through Roche-lobe overflow or winds. Some of these objects are transient stellar mass X-ray binaries
with a low mass star acting as a donor star. Study of these objects in X-rays is very interesting as they 
undergo rapid evolution in their timing and spectral properties which are strongly correlated to each other. 
There are a large number of articles by several groups (see, e.g., McClintock \& Remillard, 2006; 
Belloni et al. 2005; Nandi et al. 2012; 
Debnath et al. 2013, etc.) which discuss variation of spectral and temporal properties of these 
transient black hole candidates (BHCs) during their X-ray outbursts. In general, it has been found 
that these objects show different spectral states (hard, hard-intermediate, soft-intermediate, soft, 
etc.; McClintock \& Remillard, 2006) and low and high frequency 
quasi-periodic oscillations (QPOs) in power-density spectra (PDS) in some of these spectral states 
(Remillard \& McClintock, 2006). Different branches of X-ray color-color and 
Hardness-Intensity Diagrams (HIDs; Maccarone \& Coppi, 2003; Belloni et al. 2005; 
Debnath et al. 2008, etc.) are related to different spectral states of the outburst phases. It 
has also been observed by several authors (see, Mandal \& Chakrabarti, 2010; Nandi et al., 2012; 
Debnath et al., 2013 and references therein) that these observed spectral states show hysteresis loops 
during their spectral evolutions of an entire epoch of the outburst of these transient BHCs. Simple 
model fits of accretion rates using two component advective flow (TCAF) solution of 

Chakrabarti-Titarchuk (1995, hereafter CT95) by Mandal and Chakrabarti (2010) indicated 
that indeed the accretion rates vary differently in the rising and declining states.

In the literature, there are a large number of theoretical or phenomenological models, which 
describe accretion flow dynamics around a BH. It is well known that emitted radiation 
contains both thermal and non-thermal components. 
The thermal component is multi-color blackbody type emitted the standard Keplerian disk 
(Shakura \& Sunyaev, 1973; Novikov \& Thorne, 1973) and  the other is a power-law component, 
originated from the so-called `Compton' cloud (Sunyaev \& Titarchuk, 1980; 1985). 
This component is composed of hot electrons and is cooled down by repeated Compton scatterings 
of the low energy (soft) photons. There are many speculations about the nature of this Compton 
cloud which range from it being a  magnetic corona (Galeev, Rosner \& Viana, 1979) to a hot gas 
corona over the disk (Haardt \& Maraschi, 1991; Zdziarski et al., 2003). CT95, in their TCAF solution,
considers that the CENtrifugal pressure supported BOundary Layer or CENBOL plays the role of 
the Compton cloud. This CENBOL happens to be the post-shock region of the low-angular 
momentum halo in which a standard Keplerian disk remains immersed while emitting 
soft photons. The shock in the halo forms due to the piling up of matter behind 
the centrifugal barrier of the low angular momentum accretion 
flow component having a sub-critical viscosity parameter (Chakrabarti 1990ab, hereafter C90ab, 1996). 
These shocks are found to be stable even under non-axisymmetric 
perturbations (Okuda, Teresi \& Molteni, 2007). The other component of TCAF is an optically thick 
standard (SS73) Keplerian (disk) component which is formed in flows with super-critical viscosity 
parameter (C90ab). Of course, this disk also has to pass through the inner sonic point to satisfy the 
boundary conditions on the black hole horizon (C90ab; see also, Muchotrzeb \& Paczy\'nski, 1982).     
Formation of TCAF from a single simulation (see, Giri \& Chakrabarti, 2013; Giri, Garain \& Chakrabarti, 2015, 
and references therein) show that it has a stable configuration. Also, in Mondal et al. (2014a) a self-consistent 
transonic solution of TCAF in presence of both cooling and outflows is obtained.

Recently, after the inclusion of this TCAF solution (CT95; Chakrabarti, 1997, hereafter C97) 
in HEASARC's spectral analysis software package XSPEC as an additive table model 
Debnath, Chakrabarti \& Mondal (2014, hereafter DCM14), Debnath, Mondal \& Chakrabarti (2015,
hereafter DMC15) and Mondal, Debnath \& Chakrabarti (2014b, hereafter MDC14) obtained a clearer 
picture about the accretion flow dynamics around BHCs as they find evidences of systematically
varying accretion rates of the standard disk and the halo and the shock location on a daily basis. 
From the TCAF fitted spectrum, one can obtain two accretion rates (disk and halo), shock locations
and shock strength. These parameters also give us information about the frequency of QPOs.
Transitions of various spectral states which are observed during the outburst phases of a transient BHC
can be identified by special behaviour of accretion rate ratio (ARR) and the nature of observed QPOs. 

Newly discovered MAXI~J1659-152 is an interesting black hole binary to study because it is the 
shortest orbital period BHC observed till date (Kuulkers et al., 2010; 2013). 
The source was first observed by MAXI/GSC instrument on 25th Sept. 2010 at the sky location of 
R.A. $= 16^h59^m10^s$, Dec $= -15^\circ 16'05''$ (Negoro et al., 2010). The source was simultaneously 
observed by SWIFT/BAT instrument roughly at $17^\circ$ above the Galactic plane (Mangano et al. 2010). 
Kalamkar et al. (2011) defined the source as a BHC, based on their combined 
optical and X-ray spectral study, which was initially thought to be a Gamma-Ray Burst and was named as
GRB100925A). Kuulkers et al. (2013) based on their detailed study of the X-ray intensity variation of 
observed absorption dips (Kennea et al. 2010), confirmed MAXI~J1659-152 to be a short orbital period 
black hole binary of period $= 2.414\pm0.005$~hrs. 

MAXI~J1659-152 showed X-ray flaring activity in 2010, other than low-level activity in 2011 which
continued for $\sim 9$~months. 
During this period, the source was extensively studied in multi-wave 
band, such as various X-ray (Mu\~{n}oz-Darias et al. 2011; Kalamkar et al. 2011; Yamaoka et al. 2012; 
Kuulkers et al. 2013), optical/IR (Russel et al. 2010; Kaur et al. 2012), and radio observatories 
(Miller-Jones et al. 2011; Paragi et al. 2013). van der Horst et al. (2013) made multi-band campaign 
to explore multi-wavelength properties of the source during this outburst.
Physical parameters, such as distance, disk inclination angle, masses of the source and the companion 
are estimated using various methods. The most acceptable ranges of distance and disk inclination angle 
are $5.3-8.6$~kpc and $60-80$~degree (Yamaoka et al. 2012; Kuulkers et al. 2013) respectively. Although 
Shaposhnikov et al. (2011) predicted the mass of the BH ($M_{BH}$) as $20\pm3~M_\odot$, the preferable 
range of mass of the source and companion are $3-8~M_\odot$ (Yamaoka et al. 2012) and $0.15-0.25~M_\odot$ 
(Kuulkers et al. 2013) respectively. In this paper, we use mass of the BH as $6~M_\odot$. 

We study spectral and timing properties of the source during its 2010 main outburst phase, which continued 
for $\sim 1.5$~months, using RXTE PCA archival data. Temporal properties of the BHC along with the evolution 
of QPO frequency during declining phase of the outburst are presented in Molla et al. (2015, 
hereafter Paper II).

The {\it paper} is organized in the following way: in the next Section, we briefly discuss observation
and data analysis procedures using HEASARC's HeaSoft software package. In \S 3, we present results of 
spectral analysis using TCAF {\it fits} file as an additive table model in XSPEC and variation of 
different flow parameters extracted from model fits. Here, we also compare combined DBB and PL model fitted 
spectral analysis results with that of the TCAF fitted analysis results. Finally, in \S 4, we present 
a brief discussion and make our concluding remarks.

\section{Observation and Data Analysis}

We analyze the data of $30$ observational IDs starting from the first day of RXTE PCA observation, namely, 
2010 September 28 (Modified Julian Day, i.e., MJD = 55467) to 2010 November 11 (MJD = 55508). 
Data reduction and analysis are done using HEASARC's software package HeaSoft version HEADAS 6.15 
and XSPEC version 12.8. To analyze archival data of the RXTE PCA instrument, we follow the standard 
data analysis techniques as done by Debnath et al. (2013, 2015). 

For spectral analysis, {\it Standard2} mode Science Data of PCA (FS4a*.gz) are used. Spectra are
extracted from all the layers of the PCU2 for 128 channels (without any binning/grouping the channels).
We exclude HEXTE data from our analysis, as we find strong residuals (line features) in the HEXTE 
spectra at different energies. This could be due to the fact that the `rocking' mechanism for HEXTE stopped. 
So, we restrict our spectral analysis with the PCA data for the energy range of $2.5 - 25$ keV only.
In the entire PCA data analysis, we include the dead-time correction and also the PCA breakdown 
correction (because of the leakage of propane layers of PCUs). 
The ``runpcabackest" task was used to estimate the PCA background using the latest bright-source
background model. We also incorporated the $pca\_saa\_history$ file to take care of the SAA data. 
To generate the response files, we used the ``pcarsp" task. Detailed analysis 
will be discussed in Paper II. 

The $2.5-25$ keV PCA background subtracted spectra are fitted with TCAF based model {\it fits} file and with 
combined DBB and PL model components in XSPEC. Individual flux contributions for the DBB and PL model 
components are obtained by using the convolution model `cflux' technique.
For the entire outburst, we keep hydrogen column density (N$_{H}$) fixed at 3.0$\times$10$^{21}$~atoms~cm$^{-2}$ 
(Mu\~{n}oz-Darias et al., 2011) for absorption model {\it wabs}.
We also assume a fixed $1.0$\% systematic instrumental error for the spectral 
study during entire phase of the outburst. 
After achieving the best fit based on reduced chi-square value ($\chi^2_{red} \sim 1$), `err' command 
is used to find 90\% confidence error values for the model fit parameters. 
In Appendix, Table I, we mention average values of these two errors in superscript.

For a spectral fit, using the TCAF based model, one needs to supply five model input parameters, other than 
the normalization constant. These parameters are: $i)$ black hole mass ($M_{BH}$) in solar mass ($M_\odot$) unit, 
$ii)$ sub-Keplerian rate ($\dot{m_h}$ in $\dot{M}_{Edd}$), 
$iii)$ Keplerian rate ($\dot{m_d}$ in Eddington rate $\dot{M}_{Edd}$), 
$iv)$ location of the shock ($X_s$ in Schwarzschild radius $r_g$=$2GM/c^2$), $v)$ compression ratio ($R$) 
of the shock. The model normalization value ($norm$) is 
$\frac{R_z^2}{4\pi D^2} sin(i)$, where `$R_z^2$' represents an effective area of the emitting
region (purely on dimensional ground), $D$ is the source distance in $10$~kpc 
unit and $i$ is the disk inclination angle. In order to fit a black hole spectrum with the TCAF 
model in XSPEC, we generate model {\it fits} file ({\it TCAF0.1.fits}) 
using theoretical spectra generating software by varying five input parameters in CT95 code. We then include it 
in XSPEC as a local additive model. A brief discussion of the TCAF model, its present development and a detailed 
description of the range of input parameters and generation procedure of the current version (v0.1) of the TCAF 
{\it fits} file are given in DCM14 and DMC15. For the spectral analysis with TCAF, mass of 
the black hole is frozen at $6~M_\odot$.

\section{Results}

Accretion flow dynamics of a transient BHC can be well understood by model analysis of spectral and 
temporal behaviors of the source during its outburst phase. Here, we present results of spectral analysis 
based on TCAF and compare with combined DBB and PL model fitted results. A combined DBB and PL 
model fitted spectral analysis, though fitted well throughout and in some cases 
better than TCAF, only give gross properties of the disk such as fluxes from different components. 
However, TCAF goes one step further in extracting the detailed flow parameters, such as two disk rates and shock 
properties. Furthermore, transitions of spectral states are more conspicuous in terms of the 
fitted parameters. Thus, to study accretion dynamics around BHCs, there appears to be 
certain definite advantages in fitting with TCAF solution. A shortcoming of TCAF fit with the 
current version (v0.1) is that as the spectra become softer, the fit tends to worsen, mainly indicating that 
the importance of the halo component is reducing. In our next version this would be taken care of
by self-consistently cooling down the second component in order that the flow automatically tends to have a 
single component.

All $30$ observational IDs spreaded over the entire period of the 2010 outburst 
are initially fitted with combined DBB and PL model components in XSPEC. Model fitted 
disk temperature ($T_{in}$ in keV), power-law photon index ($\Gamma$), and flux contributions from two 
types of model components are obtained. We then refitted all the spectra with the current version of our 
TCAF model and from the fit, accretion flow parameters, 
such as disk rate ($\dot{m_d}$), halo rate ($\dot{m_h}$), location of the shock ($X_s$) and 
compression ratio ($R$) are extracted.

\subsection{Spectral Data Fitted by TCAF Solution and by Combined DBB and PL model}

A combined conventional DBB and PL model fit in $2.5-25$~keV energy range RXTE PCA spectra provides 
us with a rough estimate of flux contributions originated from both thermal (from DBB) and non-thermal (from PL) 
processes around a BH. From this, we also get an idea about the evolution of the average 
temperature of the accretion disk and spectral states by monitoring variations of $T_{in}$ and $\Gamma$ factors. 
However, from the variation of the TCAF fit parameters (such as 
two types accretion rates, $\dot{m_d}$ \& $\dot{m_h}$; 
shock parameters, $X_s$ \& $R$; and derived physical parameters, (such as the accretion rate ratio (ARR), 
shock temperature $T_{shk}$, shock height $h_{shk}$, ratio between $h_{shk}$ to $X_s$), the accretion 
flow dynamics and geometry variation during the outburst phase become very evident.
In Appendix Table I, all these fitted/derived parameters are written in a tabular form with estimated errors. 

Figures 1-3 show the variation of X-ray intensities, QPO frequencies along with TCAF and combined DBB, PL model 
fitted and derived (from TCAF) parameters. In Fig. 1a, variation of the background subtracted RXTE PCA count rate 
in $2-25$~keV ($0-58$ channels) energy band with day (MJD) is shown. In Fig. 1c, variation of TCAF fitted 
total accretion rates (combined Keplerian disk and sub-Keplerian halo rates) in the $2.5-25$~keV energy band 
are shown. For comparison, combined DBB and PL model fitted total flux variation with day (MJD) 
is shown in Fig. 1b. We observe that the variation of the TCAF fitted total flow rate (Fig. 1c)
is different from the flux variations in Figs. 1(a-b) especially in early and late stages.
In Fig. 1d, variation of {\it Accretion Rate Ratio} (ARR, defined to be the ratio of sub-Keplerian 
halo rate $\dot{m_h}$ and Keplerian disk rate $\dot{m_d}$) is plotted. Observed QPO 
frequencies (of only dominating primary QPOs) are shown in Fig. 1e. 
During the entire phase of the current outburst, only three spectral classes, such as {\it hard} (HS), 
{\it hard-intermediate} (HIMS), and {\it soft-intermediate} (SIMS) are observed. 
Strangely, the soft state (SS) is not prominent and possibly missing. 
The sequence is found to be: HIMS (rising) $\rightarrow$ SIMS $\rightarrow$ HIMS (declining) 
$\rightarrow$ HS. We believe that the absence of HS in the rising phase is due to observational constraints. The 
detailed behavior of these spectral states and other reports in the literature on them are discussed in the next Section.

In Fig. 2, variation of TCAF fitted and derived shock parameters, together with combined  DBB \& PL 
model fitted results are shown. In Figs. 2(a-b), variation of DBB temperature ($T_{in}$ in keV) 
and power-law photon index ($\Gamma$) with day (MJD) are shown. In Figs. 2(c-d), TCAF fitted shock 
location ($X_s$ in $r_g$) and compression ratio ($R$) are plotted with day. In Fig. 2(e-f), variation of 
shock height ($h_{shk}$ in $r_g$) and temperature ($T_{shk}$ in $10^{10}~K$, which is the initial temperature 
of CT95 iteration process), derived from $X_s$ \& $R$ and using Eqs. 4 \& 5 respectively of DMC15, are shown. 
In Fig. 2g, ratio between shock height and location are plotted. 
In Figs. 3(a-b), variations of DBB flux from DBB \& PL model fits and Keplerian disk rate from TCAF fits 
with day (MJD) are compared. Similarly, in Figs. 3(c-d), variations of PL flux and sub-Keplerian halo rate 
from these respective models are compared. Clearly there are `some' similarities in each pair of compared 
quantities, but not totally, since the power-law flux is a function of disk rate as well.
In Figs. 4(a-c), TCAF fitted $2.5-25$~keV background subtracted PCA spectra of three different 
spectral states (selected from approximate middle of each state to get better understanding,
marked as a, b and c in Col. 1 of Appendix Table I) along with residual $\chi^2$ are shown. In Fig. 4d, 
we show unabsorbed theoretical spectra in $0.001-3950$~keV energy range, which are used to fit
observed spectra presented in Figs. 4(a-c).

\subsection{Evolution of Spectral and Temporal Properties during the Outburst}

Detailed temporal and spectral properties of this candidate during this outburst are discussed by several 
authors on the basis of X-ray variability, QPO observations, spectral results based on inbuilt XSPEC model 
fits, such as, power-law and multi-color DBB components, etc. (Mu\~{n}oz-Darias et al. 2011; 
Kalamkar et al. 2011, etc.). However, since TCAF provides us with variation of physical parameters 
from the spectral fits on a daily basis, it may be possible find a pattern to correlate with spectral classes. 
We found interesting correlation in 2010 outburst of H~1743-322 (MDC14) and in 2010-11 outburst of GX~339-4 (DMC15). 
In Figs. 1d and 1e, we see variations of ARR and QPO frequencies with time in MJD. We find that ARR, 
total flow/accretion rate ($\dot{m_d}$+$\dot{m_h}$), shock locations, compression ratios, etc. 
in conjunction with QPOs provide a better understanding on the classification of spectral states. 
This will be discussed below.

\noindent{\it (i) Hard-Intermediate State in the Rising phase:}
RXTE has started observing the source three days after its discovery. Probably the initial low-hard state of the 
source in the rising phase was missed. For the first $3$ days of our observation (from MJD = 55467.19 to 55469.09), 
source was in a hard-intermediate state with increasing thermal DBB flux (also, Keplerian disk rate; 
see, Figs. 1 \& 3). From Fig. 3d, it is seen that during this phase, TCAF fitted sub-Keplerian halo rate shows 
a rapid fall, although combined DBB \& PL model fitted non-thermal PL flux is not changed significantly.
This fact can be better understood by observing variations of ARR in Fig. 1d. The shock moved rapidly 
(from $\sim 354$ to $207~r_g$) towards the black hole with reducing shock strength and height. This third 
observed day (2010 Sept. 30, MJD = 55469.09) is the transition day from hard-intermediate to soft-intermediate 
spectral states. QPO frequency increased monotonically (from $1.607$~Hz to $2.723$~Hz) with time (day) and ARR 
values decreased rapidly (from $3.259$ to $0.590$). According to propagating oscillatory shock (POS) 
model (Chakrabarti et al. 2005, 2008, 2009; Debnath et al. 2010, 2013; used to explain monotonic evolutions 
of QPO frequencies during rising and declining phases of the outburst), QPO rises rapidly till the compression 
ratio $R$ reaches nearly around unity as the post-shock is cooled down. This is also seen in this outburst as well (Fig. 2d).

\noindent{\it (iii) Soft-Intermediate State:}

The constancy of ARR lasted till the total rate as well as non-thermal (PL) flux or halo rate started rising suddenly on 
2010 Nov. 01 (MJD = 55501.23). This phase continued for $\sim 32$~days, where sporadic QPOs are observed with very 
little changes in $ARR$, $T_{in}$, $\Gamma$, $R$, $T_{shk}$, $h_{shk}$, and $X_s$ are observed. During this phase, the  
total X-ray intensity, flux or flow rate initially increased and then decreased mainly because of similar variations 
in thermal DBB flux or Keplerian disk rate. In this phase of the outburst, non-thermal PL flux shows decreasing pattern, 
although sub-Keplerian halo rate initially decreases, and then becomes more or less constant. On the soft-intermediate 
to declining hard-intermediate transition day, a rise in ARR value due to the effect of sudden rise in non-thermal 
PL flux/halo rate is observed. On this day, QPO frequency had a maximum ($5.951$~Hz). 

\noindent{\it (vi) Hard-Intermediate State in the Declining phase:}
This state continued for the next $\sim 3$ days, starting from the SIMS-HIMS (declining) transition day. During this phase 
QPO frequency decreases rapidly from $5.951$~Hz to $2.563$~Hz. ARR is found to increase slowly with a rise in halo 
rate compared to the disk rate (see, Figs. 1 \& 3). Rapid decrease in power-law photon-index ($\Gamma$) also indicates that 
spectrum start to become harder from the day one of this state. A slow movement of the receding shock with little 
increment in the compression ratios and shock heights are observed during this phase of the outburst. 
Nov. 05, 2010 (MJD = 55504.06) is the transition day from declining hard-intermediate to hard state. Interestingly, 
on this day, ARR is locally maximum (=$0.327$) and QPO frequency starts to decrease slowly after that. Precisely 
this behavior was seen in our earlier TCAF fits on other BHCs (see, MDC14, DCM15) as well.

\noindent{\it (vii) Hard State in the Declining phase:}

The source is observed in this spectral state till the end of the observation of 2010 outburst starting from 
the transition day. In this state, ARR (from $0.327$ to $0.278$) as well as observed QPO frequencies 
(from $2.563$~Hz to $1.638$~Hz) decrease monotonically as in other objects fitted by TCAF. A slow decrease 
in PCA count rate, total (DBB+PL) fluxes and total flow (disk+halo) rates are observed with a similar decreasing 
trend in both thermal (DBB and $\dot{m_d}$) and non-thermal (PL or $\dot{m_h}$) flux/rate components. This is 
because the supply rate is dwindling after the peak outburst is over (see, Figs. 1 \& 3). A fast receding shock 
(from $\sim 103$ to $416~r_g$) with rise in compression ratio and shock height are observed. At the same time, 
during this phase of the outburst, a decrease in power-law photon index is observed, which indicates that the 
spectrum becomes harder with a clear dominance by the sub-Keplerian halo and non-thermal power-law photons. 
Shock temperature ($T_{shk}$ of initial iteration) values are found to decrease monotonically with time (day).

\section{Discussions and Concluding Remarks}

We study the evolution of spectral properties of Galactic transient black hole candidate MAXI~J1659-152 during its 
first (2010) X-ray outburst using the current version v0.1 of two component advective flow (TCAF) solution based 
model after its inclusion as a local additive table model in HEASARC's spectral analysis software package XSPEC 
(DCM14). This has been done with a model {\it fits} file using $\sim 4\times10^5$ theoretical spectra which are 
generated by varying five model input parameters (two types of accretion i.e., Keplerian disk $\dot{m_d}$, 
sub-Keplerian halo $\dot{m_h}$ rates; two types of shock parameters: location $X_s$ and compression ratio $R$; 
and the mass of the black hole $M_{BH}$) to the modified CT95 code (see, DMC15 for details). We re-fitted all 
these spectra of MAXI~J1659-152 with combined DBB and PL model components to get a rough estimate 
about the variations of the thermal (DBB) and non-thermal (PL) fluxes during the outburst and compare these with 
our TCAF fitted results (see, Figs. 1-3). In Appendix Table I, detailed results of our spectral fit 
with observed QPO frequencies are presented.

The entire period of the 2010 outburst of MAXI~J1659-152 appears to have three spectral classes: {\it hard, 
hard-intermediate,} and {\it soft-intermediate}. It did not reach the soft state. When we study variations of 
TCAF parameters in these states, we find that there is a pattern in how the rates, ARR, QPO frequency etc. 
behave. These behaviours are similar to what were reported in other sources, (see, MDC14, DMC15). Specifically, 
we see a local maximum of ARR during the transition between hard-intermediate to hard states in all these sources. 
Clearly more objects need to be fitted before any firm conclusion can be drawn.

It is interesting that unlike other sources (MDC14, DMC15), this object exhibited no soft-states during 
this outburst according to our model.  Only for two days, MJD=$55481.71$ and MJD=$55485.16$ the fluxes are higher 
(Table I of Appendix), but observation of LFQPOs and the presence of a dip on MJD=$55483.92$ in between, 
suggests that the state is not soft, but soft-intermediate. van der Horst et al. (2013), using a disk irradiation model 
called DISKIR (with number of free parameters significantly higher than TCAF with irradiation from CENBOL on the 
Keplerian disk), suggests that the object might have gone to a soft state. 
However, the photon index (see, Fig. 5 of van der Horst et al. 2013)
of those specific days showed significant error bars and thus it is uncertain if soft states were reached. 
It is also possible that the inclination angle might also have 
played a role in hardening the spectra (as discussed in another source by
Motta et al. 2010 and by explicit Monte-Carlo simulation by Ghosh et al., 2011). 

Although low frequency ($\sim 0.01-30$~Hz) QPOs are observed almost three decades ago, there is a debate 
on the origin of this temporal behavior in Fourier transformed power-density spectrum of the X-ray intensity 
variation. According to shock oscillation model (SOM) of Chakrabarti and his collaborators in mid-90s, it can occur 
when Compton cooling time scale roughly agrees with infall time scale (Molteni, Sponholtz \& Chakrabarti, 1996) 
or due to non-satisfaction of Rankine-Hugoniot conditions to form a stable shock (Ryu, Chakrabarti \& Molteni, 1997). 
Recent numerical simulations of Garain et al. (2014) also demonstrated this in presence of Comptonization.
According to SOM, the QPO frequency is inversely proportional to the infall time ($t_{infall}$) in the post-shock region.
It also has been observed that these QPO frequencies show monotonically increasing (during rising phase of the outburst) or 
decreasing (during declining phase of the outburst) nature in hard and hard-intermediate spectral states. This evolution of the 
QPO frequencies can be well fitted with the POS model, which is nothing but time varying form 
of the SOM. Movement of the shock inward could be due to rapid cooling and consequent collapse of the CENBOL (in the rising phase) 
and outward, due to the lack of cooling in the declining phase (Mondal, Chakrabarti \& Debnath, 2015).

In soft-intermediate spectral states, sporadic QPOs are observed, which may be due to appearance or disappearance 
of the oscillating component, namely, CENBOL by intrusion of strong toroidal magnetic fields. Strong sporadic jets 
are also seen in these states (e.g., Nandi, et al. 2001; Radhika \& Nandi 2014). Our current TCAF model takes care 
of the combined effects of CENBOL and the outflow. Separation of the effects of CENBOL and jets is possible from 
more detailed modeling of timing properties and will be incorporated in a later version of TCAF.

Recently, it has been shown (DMC15, MDC14) using examples of 2010-11 outburst of Galactic BHC GX 339-4 and 
2010 outburst of Galactic BHC H~1743-322 how spectral state transitions may be triggered when the relative 
ratio of the accretion rates, namely, ARR, vary in specific ways. The nature of variation of QPOs (when observed), 
shock locations, strengths etc. are also very specifically. Exactly same type of variation of the fitted 
parameters are also seen for the current source MAXI~J1659-152. 
From the observed variation of Keplerian rates in these objects, we believe that an outburst is triggered 
due to a sudden rise in viscosity and is turned off due to the reduction in viscosity 
(CT95, Ebisawa, et al. 1996; Chakrabarti, Dutta \& Pal, 2009). It is possible that this object
belongs to a category, with short orbital period, where accretion disk around the black hole is mostly 
dominated by the wind accretion compared to disk accretion. The Keplerian disk is always immersed inside a 
strong sub-Keplerian halo. So, the soft state may be difficult to achieve.
In future, we will make detailed spectral and temporal study of other such objects (for e.g., XTE~J1118+480 of 
orbital period $\sim 4.1$~hrs, Gonz\'{a}lez-Hern\'{a}ndez et al. 2013; Swift~J1753.5-0127 of orbital period 
$\sim 3.2$~hrs, Zurita et al. 2007) during their X-ray outbursts to check if flow dynamics of these sources also 
follow a similar trend. Prediction of QPO frequency from TCAF solution fitted shock parameters ($X_s$ \& $R$; DCM14), 
and comparative study with POS model solution will be published elsewhere.

\section*{Acknowledgments}

A. A. Molla acknowledges supports of DST sponsored Fast-track Young Scientist project fellowship 
and MoES sponsored Junior Research Fellowship. Mr. S. Mondal acknowledges the support of CSIR-NET scholarship.
DD acknowledges supports from project funds of DST sponsored Fast-track Young Scientist and ISRO sponsored RESPOND.

{}



\begin{figure}
\vskip 0.0cm
\epsscale{.86}
\plotone{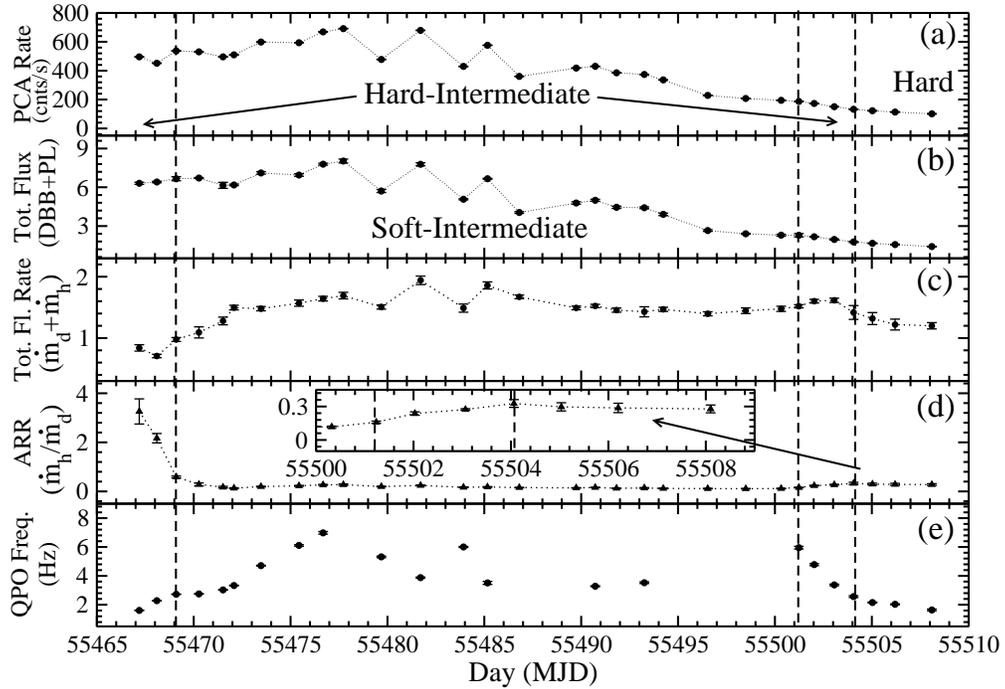}
\caption{Variation of (a) $2-25$~keV PCA count rates (cnts/sec), (b) combined disk black body (DBB) and power-law
(PL) model fitted total spectral flux in $2.5-25$~keV range (in units of $10^{-9}~ergs~cm^{-2}~s^{-1}$),
(c) TCAF model fitted total flow (accretion) rate (in $\dot{M}$$_{Edd}$; sum of Keplerian disk, $\dot{m_d}$ and
sub-Keplerian halo $\dot{m_h}$ rates) in the $2.5-25$~keV energy band, and (d) Accretion Rate Ratio (ARR; ratio
between halo and disk rates) with day (MJD) for the 2010 outburst of MAXI~J1659-152 are shown.
In the bottom panel (e), observed primary dominating QPO frequencies (in Hz) with day (MJD) are shown.
The vertical dashed lines indicate the transitions of between different spectral states.}
\label{fig1}
\end{figure}

\begin{figure}
\vskip -0.2cm
\epsscale{.86}
\plotone{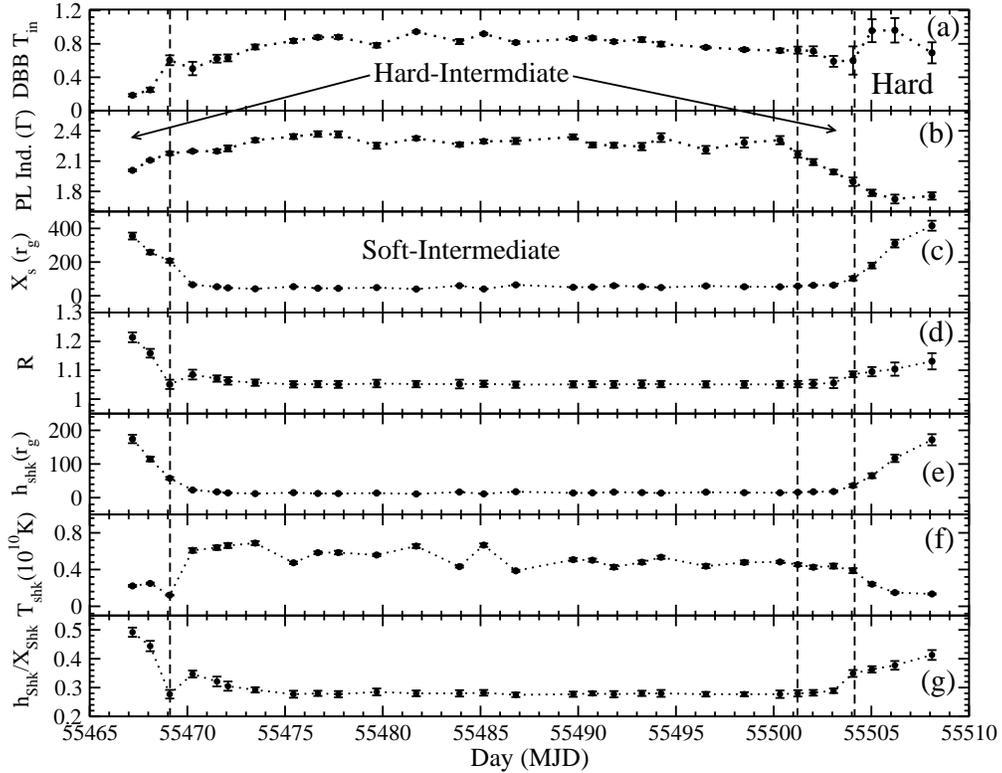}
\label{fig2}
\caption{Variation of combined DBB and PL model fitted (a) disk temperature $T_{in}$ (in keV), and (b) 
PL photon index ($\Gamma$) with day (MJD) are shown in top two panels. Variation of TCAF model fitted/derived 
shock (c) location ($X_s$ in $r_g$), (d) compression ratio ($R$), (e) temperature ($T_{shk}$ in $10^{10}$ K),
(f) height ($h_{shk}$ in $r_g$), and (g) ratio between $h_{shk}$ \& $X_s$, with day (MJD) are shown. 
The shock height, and temperature are derived from Eqs. 4 \& 5 respectively of DMC15.
}
\end{figure}

\begin{figure}
\vskip 0.8cm
\epsscale{.98}
\plotone{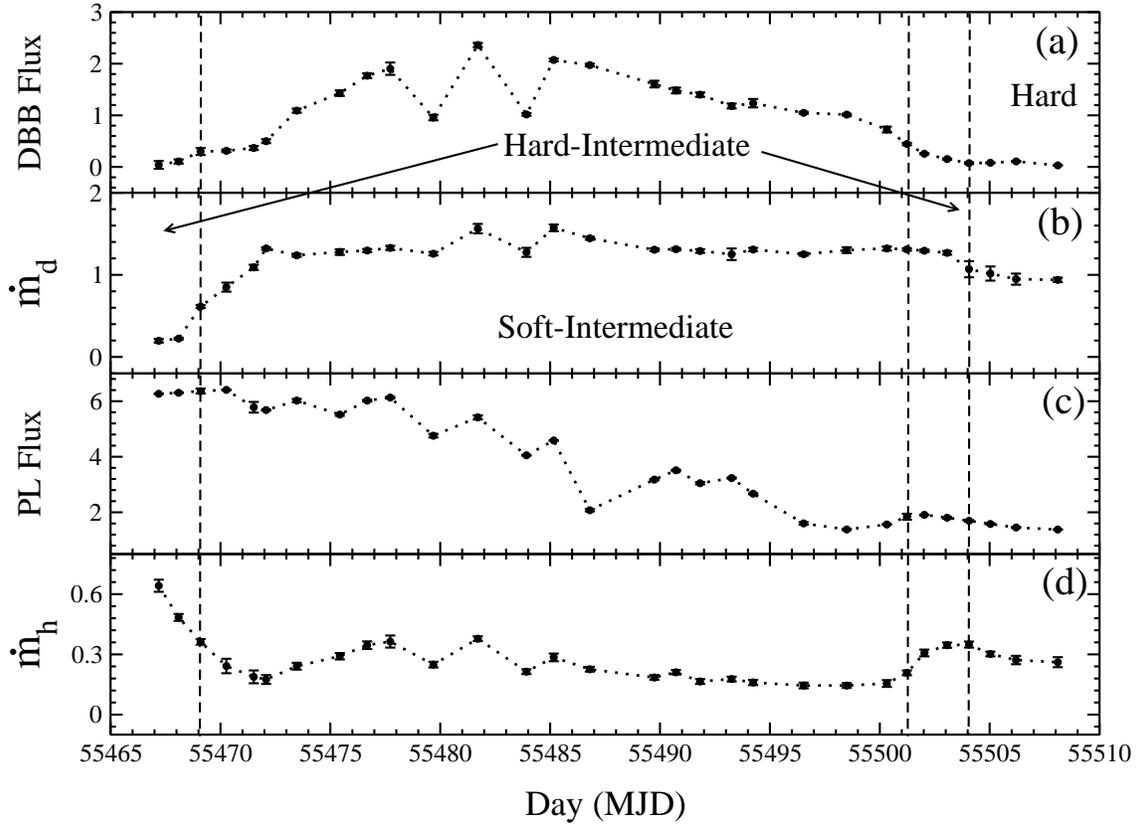}
\label{fig3}
\caption{In top panel (a), the variation of combined disk black body (DBB) and power-law (PL) model fitted DBB
spectral flux and in panel (c), the variation of PL spectral flux (both in units of $10^{-9}~ergs~cm^{-2}~s^{-1}$)
in $2.5-25$~keV energy range are shown. In panel (b), the variation of TCAF model fitted Keplerian disk rate $\dot{m_d}$
and in bottom panel (d), the variation of sub-Keplerian halo rate $\dot{m_h}$ (both in $\dot{M}$$_{Edd}$) in the
same energy band are shown.
}
\end{figure}

\begin{figure}
\centerline{
\includegraphics[scale=0.6,angle=270,width=5truecm]{fig4a.ps}
\includegraphics[scale=0.6,angle=270,width=5truecm]{fig4b.ps}
}
\centerline{
\includegraphics[scale=0.6,angle=270,width=5truecm]{fig4c.ps}
\includegraphics[scale=0.6,angle=270,width=5truecm]{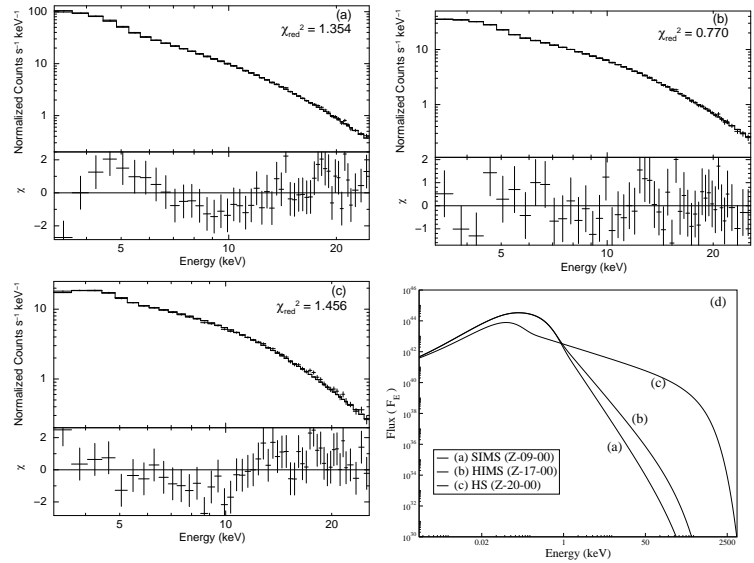}
}
\caption{TCAF model fitted $2.5-25$~keV PCA spectral flux, in units of $Normalized Counts~~s^{-1}~keV^{-1}$
with variation of $\Delta \chi$, selected from three different spectral states: soft-intermediate, hard-intermediate, 
and hard states respectively are shown in plots (a-c). The observation IDs (where Z=95118-01) and fitted values of $\chi^2_{red}$ 
are written down in the plots. In bottom-right plot (d), unabsorbed TCAF model generated spectra, which are used to fit 
these spectra are shown.
\label{fig4}}
\end{figure}

\clearpage

\begin{table}
\vskip 0.5cm
\addtolength{\tabcolsep}{-3.50pt}
\centering
\centering{\large \bf Appendix I}
\vskip 0.2cm
\centerline {2.5-25 keV Combined DBB-PL and TCAF Model Fitted Parameters with QPOs}
\vskip 0.2cm
\begin{tabular}{lcccccccccccccccc}
\hline
\hline
Obs. Id&MJD&$T_{in}$&$\Gamma$&$DBBf^\dagger$&$PLf^\dagger$&$\chi^2/DOF$&$\dot{m_d}$&$\dot{m_h}$&ARR&$X_s$&R&$h_{shk}$&$T_{shk}$&QPO$^{\dagger\dagger}$&$\chi^2/DOF$\\
 & & (keV) & & & & &($\dot{M}$$_{Edd}$)&($\dot{M}$$_{Edd}$)&&($r_g$)& &($r_g$)&($10^{10}K$)& (Hz) &   \\
 (1)&  (2)  & (3)  & (4)& (5) & (6) & (7) &  (8) & (9) & (10) & (11) & (12) & (13) & (14) & (15) & (16) \\
\hline
X-02-00&55467.19&$0.184^{0.014}$&$2.008^{0.009}$&$0.041^{0.075}$&$6.265^{0.005}$&68.1/46&$0.197^{0.022}$&$0.642^{0.030}$&$3.259^{0.513}$&$354.4^{20.4}$&$1.214^{0.017}$&$174.4^{12.5}$&$0.222^{0.010}$&$1.607^{0.009}$&47.6/35 \\
X-02-01&55468.09&$0.249^{0.025}$&$2.109^{0.009}$&$0.105^{0.040}$&$6.302^{0.014}$&64.3/46&$0.223^{0.012}$&$0.484^{0.017}$&$2.170^{0.190}$&$258.4^{12.0}$&$1.159^{0.015}$&$114.8^{6.83}$&$0.249^{0.009}$&$2.278^{0.016}$&68.4/45 \\
 & & & & & & & & & & & & & & & \\
X-02-02&55469.09&$0.603^{0.059}$&$2.176^{0.016}$&$0.299^{0.067}$&$6.364^{0.088}$&41.8/50&$0.614^{0.018}$&$0.362^{0.014}$&$0.590^{0.040}$&$207.5^{10.8}$&$1.051^{0.016}$&$57.55^{3.86}$&$0.121^{0.005}$&$2.723^{0.015}$&57.6/45 \\
X-03-00&55470.26&$0.504^{0.080}$&$2.198^{0.011}$&$0.312^{0.025}$&$6.406^{0.013}$&47.9/50&$0.850^{0.055}$&$0.242^{0.036}$&$0.285^{0.061}$&$65.20^{3.29}$&$1.085^{0.017}$&$22.62^{1.49}$&$0.609^{0.025}$&$2.749^{0.014}$&69.9/45 \\
Y-03-00&55471.51&$0.622^{0.046}$&$2.199^{0.016}$&$0.367^{0.039}$&$5.782^{0.192}$&52.1/50&$1.091^{0.032}$&$0.188^{0.032}$&$0.172^{0.034}$&$53.46^{2.89}$&$1.071^{0.011}$&$17.17^{1.11}$&$0.639^{0.024}$&$3.028^{0.019}$&51.5/45 \\
Y-05-00&55472.07&$0.632^{0.041}$&$2.226^{0.029}$&$0.500^{0.038}$&$5.682^{0.019}$&43.4/50&$1.321^{0.015}$&$0.175^{0.022}$&$0.132^{0.018}$&$46.60^{2.71}$&$1.063^{0.013}$&$14.21^{1.00}$&$0.662^{0.027}$&$3.329^{0.023}$&38.3/45 \\
Y-09-00&55473.47&$0.763^{0.028}$&$2.306^{0.019}$&$1.091^{0.040}$&$6.021^{0.058}$&38.9/50&$1.238^{0.018}$&$0.241^{0.017}$&$0.195^{0.017}$&$41.17^{1.79}$&$1.057^{0.011}$&$12.01^{0.65}$&$0.688^{0.022}$&$4.709^{0.024}$&42.1/45 \\
Y-13-00&55475.43&$0.835^{0.023}$&$2.343^{0.024}$&$1.430^{0.053}$&$5.519^{0.044}$&39.1/50&$1.278^{0.035}$&$0.291^{0.016}$&$0.228^{0.019}$&$53.73^{1.85}$&$1.051^{0.011}$&$14.91^{0.67}$&$0.474^{0.013}$&$6.108^{0.050}$&54.8/45 \\
Y-17-00&55476.67&$0.878^{0.021}$&$2.368^{0.028}$&$1.768^{0.045}$&$6.020^{0.029}$&46.7/50&$1.296^{0.019}$&$0.346^{0.019}$&$0.267^{0.018}$&$44.53^{1.16}$&$1.052^{0.011}$&$12.46^{0.46}$&$0.584^{0.014}$&$6.981^{0.087}$&49.5/45 \\
Y-19-00&55477.72&$0.881^{0.022}$&$2.365^{0.030}$&$1.903^{0.120}$&$6.128^{0.028}$&31.6/50&$1.328^{0.025}$&$0.364^{0.030}$&$0.274^{0.028}$&$43.73^{1.83}$&$1.051^{0.012}$&$12.13^{0.65}$&$0.585^{0.019}$&$ ---         $&45.6/45 \\
Y-23-00&55479.68&$0.783^{0.027}$&$2.252^{0.029}$&$0.956^{0.053}$&$4.756^{0.064}$&40.3/50&$1.258^{0.021}$&$0.249^{0.014}$&$0.198^{0.015}$&$48.07^{1.01}$&$1.054^{0.013}$&$13.68^{0.45}$&$0.559^{0.013}$&$5.312^{0.055}$&45.5/45 \\
Y-27-00&55481.71&$0.945^{0.011}$&$2.326^{0.017}$&$2.362^{0.041}$&$5.418^{0.075}$&42.5/50&$1.563^{0.057}$&$0.378^{0.012}$&$0.242^{0.017}$&$39.79^{1.70}$&$1.052^{0.011}$&$11.14^{0.59}$&$0.656^{0.021}$&$3.876^{0.041}$&81.7/45 \\
Y-30-00&55483.92&$0.827^{0.029}$&$2.264^{0.020}$&$1.018^{0.025}$&$4.053^{0.016}$&45.8/50&$1.274^{0.055}$&$0.214^{0.012}$&$0.168^{0.016}$&$59.80^{1.38}$&$1.052^{0.015}$&$16.74^{0.62}$&$0.432^{0.011}$&$5.999^{0.037}$&52.4/44 \\
Z-02-00&55485.16&$0.921^{0.011}$&$2.295^{0.018}$&$2.072^{0.027}$&$4.586^{0.011}$&44.7/50&$1.573^{0.040}$&$0.285^{0.019}$&$0.181^{0.016}$&$39.87^{1.42}$&$1.053^{0.011}$&$11.53^{0.52}$&$0.666^{0.019}$&$3.512^{0.102}$&73.5/45 \\
Z-03-00&55486.80&$0.815^{0.015}$&$2.299^{0.030}$&$1.972^{0.025}$&$2.078^{0.042}$&44.9/50&$1.446^{0.018}$&$0.226^{0.012}$&$0.156^{0.010}$&$64.60^{2.64}$&$1.050^{0.011}$&$17.76^{0.91}$&$0.386^{0.012}$&$ ---         $&70.1/45 \\
Z-06-01&55489.74&$0.864^{0.018}$&$2.339^{0.025}$&$1.605^{0.064}$&$3.179^{0.026}$&50.9/50&$1.306^{0.018}$&$0.185^{0.012}$&$0.142^{0.011}$&$50.23^{2.44}$&$1.051^{0.011}$&$13.93^{0.82}$&$0.508^{0.018}$&$ ---         $&58.3/45 \\
Z-07-00&55490.72&$0.869^{0.019}$&$2.259^{0.024}$&$1.481^{0.055}$&$3.513^{0.020}$&46.1/50&$1.312^{0.016}$&$0.211^{0.011}$&$0.161^{0.010}$&$51.64^{2.41}$&$1.052^{0.011}$&$14.45^{0.83}$&$0.502^{0.017}$&$3.283^{0.030}$&45.9/45 \\
$^a$Z-09-00&55491.82&$0.826^{0.017}$&$2.256^{0.025}$&$1.399^{0.043}$&$3.048^{0.042}$&44.7/50&$1.291^{0.022}$&$0.165^{0.012}$&$0.128^{0.011}$&$59.58^{4.82}$&$1.051^{0.012}$&$16.53^{1.53}$&$0.427^{0.022}$&$ ---         $&60.9/45 \\
Z-10-00&55493.25&$0.851^{0.027}$&$2.244^{0.037}$&$1.181^{0.048}$&$3.235^{0.020}$&64.6/50&$1.251^{0.070}$&$0.177^{0.012}$&$0.141^{0.017}$&$53.94^{2.94}$&$1.052^{0.013}$&$15.09^{1.01}$&$0.480^{0.019}$&$ ---         $&82.7/45 \\
Z-11-00&55494.23&$0.797^{0.027}$&$2.333^{0.044}$&$1.235^{0.081}$&$2.669^{0.017}$&54.3/50&$1.308^{0.021}$&$0.160^{0.013}$&$0.122^{0.012}$&$48.60^{2.02}$&$1.052^{0.011}$&$13.60^{0.71}$&$0.534^{0.017}$&$ ---         $&46.4/45 \\
Z-13-00&55496.53&$0.758^{0.015}$&$2.212^{0.038}$&$1.049^{0.019}$&$1.602^{0.046}$&45.0/50&$1.251^{0.018}$&$0.145^{0.014}$&$0.116^{0.013}$&$58.24^{4.47}$&$1.051^{0.011}$&$16.16^{1.41}$&$0.437^{0.021}$&$ ---         $&67.1/45 \\
Z-15-00&55498.49&$0.731^{0.018}$&$2.283^{0.048}$&$1.014^{0.017}$&$1.385^{0.025}$&51.5/50&$1.301^{0.035}$&$0.145^{0.012}$&$0.111^{0.012}$&$53.33^{3.57}$&$1.051^{0.012}$&$14.79^{1.16}$&$0.478^{0.021}$&$ ---         $&60.7/45 \\
Z-16-00&55500.31&$0.719^{0.023}$&$2.307^{0.041}$&$0.725^{0.056}$&$1.564^{0.014}$&58.3/50&$1.321^{0.025}$&$0.155^{0.017}$&$0.117^{0.015}$&$52.80^{1.74}$&$1.051^{0.012}$&$14.65^{0.65}$&$0.483^{0.013}$&$ ---         $&57.8/45 \\
 & & & & & & & & & & & & & & & \\
Z-16-01&55501.23&$0.725^{0.041}$&$2.167^{0.033}$&$0.444^{0.023}$&$1.842^{0.107}$&52.1/50&$1.310^{0.015}$&$0.208^{0.012}$&$0.159^{0.011}$&$57.07^{3.18}$&$1.052^{0.011}$&$15.97^{1.06}$&$0.454^{0.017}$&$5.951^{0.091}$&41.2/45 \\
$^b$Z-17-00&55502.02&$0.713^{0.057}$&$2.090^{0.030}$&$0.256^{0.011}$&$1.909^{0.008}$&42.5/50&$1.294^{0.014}$&$0.307^{0.017}$&$0.237^{0.015}$&$61.79^{3.93}$&$1.053^{0.014}$&$17.44^{1.34}$&$0.426^{0.019}$&$4.779^{0.070}$&34.7/45 \\
Z-17-01&55503.06&$0.590^{0.065}$&$1.992^{0.023}$&$0.152^{0.013}$&$1.806^{0.033}$&41.4/50&$1.269^{0.022}$&$0.346^{0.013}$&$0.273^{0.015}$&$63.00^{5.39}$&$1.056^{0.018}$&$18.23^{1.87}$&$0.438^{0.026}$&$3.371^{0.056}$&34.3/45 \\
 & & & & & & & & & & & & & & & \\
Z-18-00&55504.06&$0.600^{0.168}$&$1.895^{0.042}$&$0.073^{0.016}$&$1.695^{0.031}$&49.6/50&$1.068^{0.098}$&$0.349^{0.015}$&$0.327^{0.044}$&$102.8^{12.2}$&$1.086^{0.010}$&$35.83^{4.58}$&$0.388^{0.027}$&$2.563^{0.060}$&53.6/45 \\
Z-19-00&55505.03&$0.958^{0.137}$&$1.786^{0.032}$&$0.081^{0.008}$&$1.583^{0.023}$&46.5/50&$1.016^{0.086}$&$0.302^{0.012}$&$0.297^{0.037}$&$179.1^{17.1}$&$1.095^{0.016}$&$65.08^{7.17}$&$0.241^{0.015}$&$2.154^{0.034}$&52.8/45 \\
$^c$Z-20-00&55506.20&$0.962^{0.146}$&$1.724^{0.044}$&$0.107^{0.011}$&$1.456^{0.018}$&46.5/50&$0.948^{0.068}$&$0.272^{0.021}$&$0.287^{0.042}$&$310.0^{23.2}$&$1.104^{0.023}$&$116.9^{11.2}$&$0.149^{0.009}$&$2.028^{0.033}$&65.4/45 \\
Z-21-00&55508.09&$0.692^{0.126}$&$1.755^{0.036}$&$0.031^{0.009}$&$1.381^{0.011}$&43.8/50&$0.939^{0.027}$&$0.261^{0.025}$&$0.278^{0.035}$&$416.2^{29.7}$&$1.131^{0.028}$&$171.9^{16.5}$&$0.134^{0.008}$&$1.638^{0.056}$&60.4/45 \\
\hline
\end{tabular}
\noindent{
\leftline {Here X=95358-01, Y=95108-01, and Z=95118-01 mean the initial part of the observation Ids, and (a-c) mark TCAF model fitted results for three different states,}
\leftline {presented in Fig. 4. Intermediate void space mark state transitions from HIMS$\rightarrow$SIMS, SIMS$\rightarrow$HIMS, and HIMS$\rightarrow$HS respectively.}
\leftline {$T_{in}$, and $\Gamma$ values indicate combined DBB and PL model fitted multi-color disk black body temperatures in keV and power-law photon indices respectively. }
\leftline {$^\dagger$ DBBf, PLf represent combined DBB and PL model fitted 2.5-25~keV fluxes for DBB and PL model components respectively in units of $10^{-9}~ergs~cm^{-2}~s^{-1}$. }
\leftline {$\dot{m_h}$, and $\dot{m_d}$ represent TCAF fitted sub-Keplerian (halo) and Keplerian (disk) rates in Eddington rate respectively. $X_s$ (in Schwarzchild radius $r_g$), and}
\leftline {$R$ are the model fitted shock location and compression ratio values respectively. $h_{shk}$ (in $r_g$) and $T_{shk}$ (in $10^{10}~K$) are the shock height and temperature values}
\leftline {derived from Eqs. 4 \& 5 of DMC15 respectively. $^{\dagger\dagger}$ Here, frequencies of the principal QPO in Hz are presented. DOF means degrees of freedom of the model fit.}
\leftline {Note: average values of 90\% confidence $\pm$ error values obtained using `err' task in XSPEC, are mention as superscript of the spectral fitted/derived parameters.}
}
\end{table}

\end{document}